# Business Process Re-engineering in Supply Chains
## Examining the case of the expanding Halal industry

Mohammed Belkhatir[1], Shalini Bala
*School of Information Technology, Monash University Sunway Campus, 46150 Petaling Jaya, Malaysia*
[1]*belkhatir.mohammed@infotech.monash.edu*

Noureddine Belkhatir
*LIG-IMAG Laboratory, CNRS, France*
*noureddine.belkhatir@iut2.upmf-grenoble.fr*

Abstract:   Due to several issues arising in the rapidly-expanding Halal industry, among them the production of non-genuine or contaminated products and meats, there is a need to develop effective solutions for ensuring authenticity and quality. This paper proposes the specification of a formalized supply chain framework for the production and monitoring of food and products. The latter enforces high-level quality of automated monitoring as well as shorter production cycles through enhanced coordination between the actors and organizations involved. Our proposal is guided by business process support to ensure quality and efficiency of product development and delivery. It moreover meets the requirements of industrial standards by adopting the Capability Maturity Model Integration's highest process maturity level through establishing quantitative process-improvement objectives, proposing the integrated support of engineering processes, enforcing synchronization and coordination, drastic monitoring and exception handling. We then delve into some of the important technologies from the implementation point-of-view and align it with the formalized Halal framework. An Information Technology support instantiation is proposed leading to a use case scenario with technology identification.

## 1 INTRODUCTION

Halal is an Arabic term which means 'permissible' or sanctioned by Islamic law. Today, Halal certified products are products that abide by the Islamic law and include hygiene, sanitation and safety qualities (SIRIM, 2004) which are important quality aspects of food. The consumers of theses products include Muslims and non-Muslims that hail from many countries such as Asian countries, Middle East countries, America, Canada, United Kingdom, Africa and Europe. According to (AAP, 2007), the average global Halal food trade or market is estimated at 560 billion US dollars a year.

For manufacturers and service providers, the Halal standard means a larger market share to be tapped, which will simultaneously bring in more profits. As quality is associated with the standard, it is of utmost importance that the products and meat sold comply with it, but many times it has been found that this is not the case in several countries (Halal Journal, 2006).

In the USA, Europe (including the UK), Canada and Singapore, manufacturers and distribution outlets such as food sellers are using fake certificates or labels on products and meat; e.g. (IFANCA, 2002), (MUIS, 2000)… Also, there is cross-contamination in the production of Halal and non-Halal food at the manufacturers. As far as animal welfare is concerned, there is clumsy slaughtering with dull knives (PETA, 2002) which does not comply with the MS1500:2004 standard accepted by the United Nations. Animals are moreover reported to be starved before the slaughtering according to slaughterhouse officials as well as cruelly transported by traders such as from the supplier to the abattoir (PETA, 2002). However, 'animals subjected to cruelties in their breeding, transport, slaughter, or in general welfare, meat from them is considered impure', in other words not Halal (MUIS, 2000). All of these issues have caused the consumers to be concerned about the quality of the Halal food and products they purchase and eat on a daily basis.

Today, authorized certification organizations have been set up in many countries to monitor and inspect the abattoirs, manufacturers and distribution outlets in their handling of the animals and products, and to issue the Halal certificates to them once the inspections are approved. These certification organizations have been successful to a certain extent in reducing the quality problem associated with the products delivered, but there is room for improvement. Currently, these organizations are carrying out their inspections manually. Some require companies to manually submit the application to their offices and liaise with third party organizations/labs to test the products, which delay the certification process and ultimately delay the genuine products from reaching the consumers' hands quickly.

Hence, this paper proposes a framework to better counter the non-genuine products sold in the market by introducing the automation of inspections, high-level quality of monitoring, shorter certification process, shorter production cycles in the supply chain, enhanced coordination between the actors or organizations in the industry and drastic exception handling to manage and control faulty products. Process support should guide our proposal for the framework to accelerate quality product development and delivery. However, development processes must also meet the requirements of industrial standards by adopting for example the highest process maturity level of the Capability Maturity Model Integration (Chrissis et al, 2006). The framework in this paper aims to adopt some standard practices and accomplish its proposal in the following ways: i) establishing quantitative process-improvement objectives, ii) integrated support of engineering processes, iii) synchronization/coordination, iv) monitoring, v) exception handling.

Thus, technology is required to accomplish the above and to enable a better control and monitoring of the non-genuine products sold in the market, which is presented in the following sections. In Section 2, the supply chain framework is presented along with its actors and business processes. Next is section 3 which presents the Information Technology architecture supporting the framework. In Section 4 there is focus on related works and Section 5 concludes the paper.

## 2 THE SUPPLY CHAIN FRAMEWORK

The Supply Chain Framework has been modeled to identify the processes that need to be improved and re-engineered to better counter the problem of non genuine products sold in the market.
We first provide an outline of the framework and explain its aim and objectives. We moreover list the supply chain organizations and detail their role in regards to the various Halal processes. Business processes and their quantitative process-improvement objectives are then listed.

### 2.1 Outline

The solution designed in Figure 1 is the general model of the Halal Supply Chain Framework which has several objectives to fulfill in order to achieve this aim, as follows:

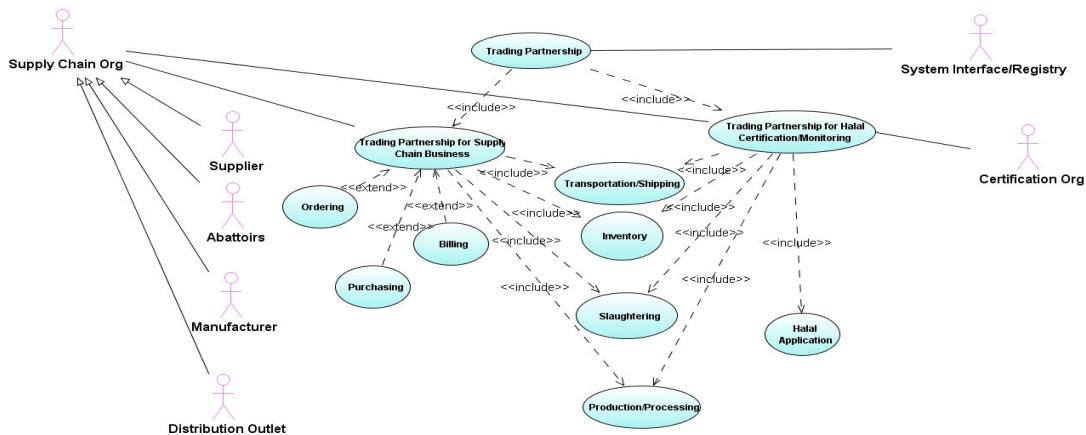

Figure 1: General Framework for the Halal Supply Chain

- Ensure continuous improvement of the performance of Halal processes via incremental and innovative technological improvements
- Enhance synchronization/coordination between the actors or organizations in the industry
- Achieve high-level quality of monitoring
- Shorten the certification process
- Shorten the production cycles in the supply chain

## 2.2 Actors

The different actors that have been identified in this framework are the Supply Chain Organizations, the Certification Organizations and the System Interface/Registry. The Supply Chain Organizations include: the suppliers/breeders who supply the animals, the abattoirs which buy these animals and slaughter them, the manufacturers who purchase the slaughtered meat and processes it for production and the distribution outlets including retailers, food sellers and butchers selling the products.

The system Interface/Registry enables the synchronization/coordination of the organizations in the supply chain so that they can communicate with each other and form an electronic trading partnership in order to better control the non genuine products. The Certification Organization plays a very important role in the framework that is to certify and monitor the organizations to ensure the authenticity of the products.

## 2.3 Business Processes

The first business process involved is the main 'trading partnership'. The framework contributes to the Halal industry by illustrating that a trading partnership is necessary before any kind of supply chain business or monitoring can be executed between the actors in this diagram. There are two types of partnership processes that can be formed between the actors.

First, the 'trading partnership for supply chain business' involves the partnership among the supplier/breeder, abattoir, manufacturer and distribution outlet in order to ensure fast production and counter the non-genuine products sold in the market. This is accomplished by accelerating the exchange of information in the processes beginning with the ordering of products, then their purchase, transportation/shipping, storing in the inventory, slaughtering, production/processing and ending with the billing procedure.

Second is the 'trading partnership for certification/monitoring' which involves the partnership between the certification organization and the supplier, abattoir, manufacturer and distribution outlet to monitor the Halal problem. It also covers the partnership between the supplier/breeder, abattoir, manufacturer and distribution outlet to increase the responsibility, cooperation and communication of each of these organizations.

Let us note that among the processes mentioned, the transportation/shipping, inventory, slaughtering and production/processing carry an important role since they are part of both trading partnerships. Indeed, these core processes are meant to accelerate product supply and are also necessary to monitor the several issues identified in the supply chain.

The common quantitative process-improvement objectives for these core processes are as follows:

- Increase the number of electronic partnerships, technical communications and documents between the certification organization with the supply chain organizations as well as among the supply chain organizations, to enhance their coordination.
- Reduce the time it takes for the certification organizations to inspect or exchange information.

The specific quantitative process-improvement objectives for the core processes are related to the rapid decrease of the number of:

- 'injured' or 'unhealthy' animals certified by veterinarians in their health reports for transported animals as far as **transportation and shipping** are concerned.
- 'starved' or 'unhealthy' animals certified by veterinarians in their health reports which are reported just before the animals are slaughtered when handling the **inventory**.
- animals that are not **slaughtered** according to the MS1500:2004 standard .
- defect products either cross-contaminated or with fake labels attributed by manufacturers during **production and processing**.
- both fake certificates put up by distribution outlets such as food sellers and defect products that are sold with fake labels at the **distribution** stage.

# 3 INFORMATION TECHNOLOGY SUPPORT INSTANTIATION

The Information Technology framework is designed to model the electronic partnerships in the Halal industry and highlight the role of the system registry which enables these partnerships. It furthermore illustrates how all the components come together to support the quantitative process-improvement objectives. This framework is based on the e-business standard technology and in particular the ebXML scenario 2 example framework (Webber, 2004). The e-business standard technology both automates and standardizes business processes so that different organizations with different systems and business processes can exchange information or documents quickly. In addition to this, this technology enables the organizations in the industry to monitor and control electronically the several issues identified, that is from a distance and in a secure manner. Also, it makes it possible to produce and certify the products fast in order to reach the consumers' hands quickly. The facility to form electronic partnerships among these organizations also enables each one of them to take the responsibility of monitoring at every stage of the supply chain. As a matter of fact, this role is no longer specific to the certification organization alone.

The partnerships, system registry components and solution to support the quantitative process-improvement objectives are presented in detail below.

## 3.1 Partnerships

In this framework, the supply chain organizations such as the suppliers/breeders, abattoirs, manufacturers and distribution outlets as well as the certification organizations have to first build a system interface that is compliant with the system registry, and then register with the latter before establishing an electronic partnership with each other. These partnerships can be formed between organizations known to each other, or new organizations intending to form partnerships with one another. Once these organizations have formed a partnership, they can proceed to communicate and liaise extensively with each other. The dynamic aspect of this process which allows new organizations to find one other in the system registry enables a vast expansion of partnerships in the industry, with unlimited geographical boundary, to better control the highlighted issues.

## 3.2 System Registry and its Components

The system registry is the system that enables different organizations in the industry to discover one another and register to form electronic partnerships. The components stored in this registry include the Library and Models, List of Scenarios, Collaboration Protocol Profiles (CPP), Messaging Constraints and Security Constraints.

The Library and Models component is based on the Business Process and Information Meta Model concept identified in a specific e-business standard technology. It consists of the definition of business processes as well as reusable core components that reflect common business semantics, XML vocabularies and actual message structures defined. The latter are reused by organizations in the industry to ensure sound communication of the standard terms and formats.

The modeling of the processes and scenarios is executed via the UN/CEFACT Modeling Methodology (UMM) which is based on the Unified Modeling Language (UML). The modeling includes the class diagram modeling of the core processes. The general model of the 'Formalized Supply Chain Framework' serves as the foundation to form the modeling of the class diagrams.

The information from the Library and Models contributes to the formation of the Collaboration Protocol Profile (CPP) which is registered in the system registry. The CPP contains information such as the industry classification, supported business processes, requirements for interface and messaging and service information for contact purposes. Other registered organizations in the industry can then discover these CPPs and find out the supported processes and scenarios and decide to become partners with each other. The agreement for a partnership is called the Collaboration Protocol Agreement (CPA). Included in the CPPs and CPAs, are the messaging and security constraint details.

## 3.3 Supporting the Quantitative Process-Improvement Objectives

The first quantitative process-improvement objective of increasing the number of electronic partnerships, technical communications and documents among the organizations in the industry, can be achieved through the electronic partnerships and System Registry components. As more and more organizations register in the system registry and become partners with one another, the number of CPPs and CPAs increase, which indicates that more organizations are able to assist the certification organization to monitor the highlighted issues in the supply chain.

This monitoring will be executed via XML documents that are exchanged between the supply chain organizations and the certification organization as well as among themselves. The increase of XML documents designed and exchanged would indicate extensive communication, cooperation and coordination between these actors.

By communicating via XML, the second quantitative process-improvement objective can be supported which is reducing the time it takes for the certification organizations to inspect or exchange information. The Partnerships and System Registry components, together, provide real-time information access, faster information exchange and query resolution as well as improved information flow. They furthermore enable mutual use of the information that has been exchanged between the organizations in the supply chain. The monitoring is assisted as certain aspects of the inspections are automated and executed from a distance with speed and security. For instance, the health reports by veterinarians who check the animals after their transportation from the supplier/breeder to the abattoir can be transmitted to the certification organization electronically, eliminating the need for physical inspectors to look through the health reports. Also, the submission of applications and documents to the certification organization, and the liaising between the latter and third party organizations/labs to test the products can also be achieved via XML technology. The fast exchange of information in turn assists to shorten the certification process and the production cycle.

The above stated solution also supports the specific quantitative process-improvement objectives of rapidly decreasing the number of 'unhealthy' animals in the transportation and inventory process. This is because after the animals are transported and before they are slaughtered, health reports by veterinarians are transmitted to the certification organization and these reports would be recording the number of animals that are certified 'unhealthy'. In this way, the number of 'unhealthy' animals can be monitored at a closer interval of inspection due to the utilization of technology and can be decreased at a faster pace. With the Halal Partnerships and System Registry, the transportation process can also be inspected by the supplier/breeder and abattoir, not only by the certification organization. This is because either the supplier/breeder will engage the transportation service or the abattoir will engage the service for transporting the animals. Since the process involves two different organizations in the supply chain, these can help the certification organization by monitoring each other via the e-business technology and documents, and inform the certification organization whether there is a problem in the transportation service such as lack of room which can harm the animals.

As for the rest of the specific quantitative process-improvement objective in the slaughtering, production/processing and distribution process, the certification organization will need to physically inspect these processes to ensure that all requirements are met, but unlike before, all the inspection reports will be transmitted via the e-business system immediately after the inspection to speed up the resolution of issues encountered.

## 4 RELATED WORKS

According to Kok (Kok, 2003) and Kotinurmi et al. (Kotinumi et al, 2003), ebXML and RosettaNet (RosettaNet, 2007) are among prominent standards that provide key business and technology benefits for e-business integration (Webber, 2004) and have a high level of general adoption across the globe. Both standards focus on integrating different systems and business processes in several organizations to execute business easily with each other. Nevertheless, the two standards differ from one another as the RosettaNet standard specifically targets the high-tech industry whereas ebXML does not target any industry. In addition to the ebXML and RosettaNet standards, OAGIS and xCBL are also known to be among the prominent XML-based e-business standards (Numilaakso et al, 2006) that have high likelihood of general adoption due to their suitability for industrial procurement, design, production or distribution. xCBL is however migrating in phases to UBL which has now taken a prominent role.

To address the issues in the Halal supply chain, we have selected the ebXML standard technology. Following are the reasons discussed.

OAGIS and UBL (since xCBL is migrating in phases to UBL, we will restrict ourselves to discussing the latter) are non-proprietary and open standards, which are major advantages. Furthermore, they have been implemented in food industries where they provide reasonable cost and support. They also provide processes which are particularly suited for the Halal industry such as the certification of origin process which can assist in verifying imported products. However, these technologies are not complete and have therefore to be instantiated along with other e-business technologies such as ebXML messaging, registry and security mechanisms.

The RosettaNet technology provides many advantages as ebXML, but is too focused on the Electronics industry and its processes and documents would require extensive modifications to suit the Halal scenario. Thus, it is concluded that the ebXML technology would be selected to address the highlighted issues in the Halal supply chain. It is furthermore deemed not necessary for ebXML to work in combination with other technologies as it has enough facilities, materials and support to work on its own.

## 5 CONCLUSION

This paper has discussed various opportunities and issues related to Halal industry and Halal product process. Even though a lot of work is going on in the business world related to Halal industry in general, not much work has been done to formalize the business processes.

The overall contribution of this paper is a process-oriented framework for managing the production of Halal products, which in particular addresses the issue of the production and commercialization of non-genuine or faulty products. We have identified the main activities at the core of the framework as well as the data and control flows linking them.

We have moreover proposed the technological solutions for its instantiation and in particular the XML-based data flow allowing the unified description of documents through a generic formalism. After the study of the related works on process engine, we have directed our choice towards the use of open-source software in order to instantiate the proposed framework and to make the highlighted processes enactable and executable.

In future works, we will work on the modeling aspects, i.e. the definition of a language allowing process specification and characterization. This language of high-level abstraction shall be independent from any particular implementation architecture. We will carry on a large experimental study based on a real-world scenario, for which we will study the feasibility of its deployment in a distributed environment.